\documentclass[preprint, aps, showpacs, floatfix]{revtex4-1}

\usepackage{graphicx}
\usepackage{amsmath,amssymb}
\usepackage[usenames,dvipsnames]{color}
\usepackage[colorlinks=true, citecolor=blue, linkcolor=WildStrawberry]{hyperref}
\usepackage{epstopdf}

\newcommand{\beq}{\begin{equation}}
\newcommand{\eeq}{\end{equation}}
\newcommand{\bdm}{\begin{displaymath}}
\newcommand{\edm}{\end{displaymath}}

\DeclareFontFamily{OT1}{pzc}{}
\DeclareFontShape{OT1}{pzc}{m}{it}{<-> s * [1.10] pzcmi7t}{}
\DeclareMathAlphabet{\mathpzc}{OT1}{pzc}{m}{it}

\graphicspath{{./plots/}}
\begin{document}

\title{Real-time earthquake warning for astronomical observatories}

\author{Michael Coughlin}
\affiliation{Department of Physics, Harvard University, Cambridge, MA 02138, USA}
\email{coughlin@physics.harvard.edu}

\author{Christopher Stubbs}
\affiliation{Department of Physics, Harvard University, Cambridge, MA 02138, USA\\
Department of Astronomy, Harvard University, Cambridge MA 02138, USA}

\author{Sergio Barrientos}
\affiliation{Centro Sismológico Nacional de la Universidad de Chile, Santiago, Chile}

\author{Chuck Claver}
\affiliation{LSST Observatory, Tucson, AZ, 85712}

\author{Jan Harms}
\affiliation{INFN, Sezione di Firenze, Sesto Fiorentino, 50019, Italy}

\author{Christopher Smith}
\affiliation{Cerro Tololo Inter-American Observatory, La Serena, Chile}

\author{Michael Warner}
\affiliation{Cerro Tololo Inter-American Observatory, La Serena, Chile}

\begin{abstract}

Early earthquake warning is a rapidly developing capability that has significant ramifications for many fields, including astronomical observatories. In this work, we describe the susceptibility of astronomical facilities to seismic events, including large telescopes as well as second-generation ground-based gravitational-wave interferometers. We describe the potential warning times for observatories from current seismic networks and propose locations for future seismometers to maximize warning times.
\end{abstract}

\maketitle

\section{Introduction}
\label{sec:Intro}


Early earthquake warning (EEW) is a rapidly developing field, which includes the efforts of California systems such as Berkeley's ElarmS and Stanford's Quake-Catcher Network, Japan's Earthquake Early Warning System, and systems in other countries, including Taiwan, Mexico, Turkey, and Romania \cite{Al2012,KuAl2013a,KuAl2013b,KuHe2014,CoLa2009a,CoLa2009b,BoAl2014,HoKa2008}. All EEW systems seek to rapidly detect and characterize earthquakes to give warning to people before the most severe shaking occurs. In general, longer warning time allows more protective action to be taken. 

Earthquakes are a potental problem for large-scale scientific experiments. They are of course dangerous as well for the personnel inside of the facilities, although in this paper, we will focus on the facilities themselves. There are three potential consequences of seismic events to scientific facilities. The first is physical damage to components and systems. Damage has economic consequences due to the procurement of replacement parts plus the cost of down time. The second is loss of operating time, which has the consequence of not taking data and therefore a loss of efficiency. The third is loss of information due to damaged data storage systems and the cost is replacing the data and information. Our goal is to maximize the protection of these facilities to minimize the losses to instruments and minimize downtime. Examples of recent experiments that have been affected by earthquakes are astronomical observatories, such as meter-class telescopes and gravitational-wave interferometers. In this study, we will explore the potential benefits of EEW for these two types of astronomical observatories. Due to design differences, these two types of facilities represent the opposite ends of the spectrum of sensitivity to earthquakes. Telescopes are primarily susceptible to large, regional earthquakes which can potentially damage the instruments, while gravitational-wave observatories are sensitive to earthquakes that occur around the world, creating ground motion that makes it difficult to take data.

In the case of telescopes, the predominant concern is the potential for large ground motions which might damage the telescope drives, instruments, enclosures or the mirrors. The telescope facilities typically have extensive performance requirements related to their ability to either remain operational during or simply survive the passing of seismic waves. The requirements are broken up into two types. ``Operational'' earthquakes are relativity small magnitude earthquakes, and modern telescope systems are designed to withstand them with no structural yielding or loss of function, with expected repair times on the order of days (not including re-verification and testing). ``Survival earthquakes'' are large earthquakes, and systems must be designed to not fail (exceed ultimate stress), especially if a failure could lead to loss of life, damage exceeding millions of dollars, or repair times on the order of years.  Thus, any significant damage to an instrument, filters, or wholesale dropping of jetsam onto the primary, second, or tertiary mirror are to be avoided. The performance requirements are created based on probabilistic ground motion levels for the sites on which the telescopes are situated, depending on the occurrence of regional earthquakes \cite{KaAn2008}. The Large Synoptic Survey Telescope (LSST), in particular, uses estimated survival levels for earthquakes that have a 10\% probability of occurring within a 30 year lifetime \cite{Ne2012}. These analyses informed the LSST seismic design requirements \cite{NeWa2012}. LSST has specifications to remain operational after sustaining accelerations of 3.8\,g transverse, 1.21\,g along +Z-axis, and 1.86 g along -Z-axis, while survival earthquake levels are 5.7\,g transverse, 2.04\,g along +Z-axis, 2.45\,g along -Z-axis.

A recent example of significant telescope downtime was from the magnitude 6.7 Hawaiian earthquake (with a 6.1 aftershock) on October 16, 2006. The epicenter was 58\,km away from the Keck telescopes, at a depth of 29\,km. After the earthquake occurred, it took more than a month for Subaru, Keck 1, Keck 2, CFHT, and Gemini to return to full operations \cite{MKO2007}. Although the domes and shutters were largely spared from damage, on Keck I the azimuth drive track and encoder surface was damaged by significant translation of the telescope, which was about 12\,mm. There was also significant damage to individual instruments.  A more recent example is the Arecibo Observatory 305 meter radio telescope in Puerto Rico which has cables that were seriously damaged from a 6.4 magnitude earthquake in January 2014. The earthquake was 78\,km from the telescope, at a depth of 20\,km. There are also examples of large earthquakes that did not generate damage. The April 1, 2014, 8.2 magnitude earthquake near Iquique, Chile did not damage the Chilean telescopes. The earthquake was 1047\,km from the Magellan telescope, at a depth of 25\,km. The Lick observatory was not damaged by the August 2014 Napa Valley Earthquake. The earthquake was 115\,km from the telescope, at a depth of 11\,km. This earthquake has been thoroughly analyzed in a recent series of articles, which show that another large earthquake is likely to occur in the same region in the near future \cite{Bur2014,HaHe2014,ScAs2014}. It is clear from recent history that damage to telescopes will depend significantly on the particular situation, specifically distance, magnitude, and the particular telescope in question. Heuristically, these examples suggest that earthquakes closer than 100\,km are required to do the most damage, and we show why this is likely to be the case in section~\ref{sec:hazard}.

The Laser Interferometer Gravitational-wave Observatory (LIGO) \cite{aLIGO}, Virgo \cite{adVIRGO}, and GEO600 \cite{Gr2010} detectors are part of a network of gravitational-wave interferometers seeking to make the first direct observations of gravitational waves. These detectors are designed to limit their sensitivity to environmental effects, such as from local magnetic fields or ground motion, but they are still susceptible to a variety of instrumental and environmental noise sources that decrease their detection sensitivity \cite{S6DetectorChar}. Environmental noise can couple into the interferometer through mechanical vibration or because of electromagnetic influence. When an interferometer is running with low noise, data are recorded in what it is called ``science time'' and the interferometer is said to be ``locked.'' 

Seismic motion from human activity near the sites, from wind, and from ocean waves are among the most common sources of disturbances that affect the detector. Earthquakes, due to the ground motion they induce, can destabilize the gravitational-wave detectors. The predominant concern is the difficulty of staying locked during the presence of ground motion from teleseismic signals from around the world \cite{MaFa2012}. The second-generation gravitational-wave detectors contain upgraded seismic isolation systems (see \cite{AbAd2002,StAb2009} and references therein). The interferometer optics use seven stages of isolation for the core optics, two internal, one external, and four passive (pendulums). Both the first external stage and internal stages use position and inertial sensors along with actuators to limit their platforms' motion. Despite this sophisticated system, the optics are still affected by the ground motion during earthquakes. Another potential concern is simultaneous arrival of transient seismic disturbances that masquerade as false gravitational-wave signals. 

During the last LIGO science run, large amplitude earthquakes from around the world would typically cause the detectors to fall out of lock. Not only were the data around the time of the earthquake not useful for gravitational-wave detection, but it would also take hours of dead time for the detectors to return to the locked state. Although there have not yet been detailed studies of this phenomenon (currently underway to characterize the effects on the advanced detectors), we can perform a simple back-of-the-envelope calculation. 
During the last data collection run, the Hanford and Livingston detectors took 245.6 and 225.2 days of science time respectively \cite{S6Lowmass}. Taking the earthquakes with magnitudes greater than 6.0 that occured during that time, loss of interferometer lock occurred 139 times for Hanford and 127 times for Livingston within an hour of the expected surface wave arrivals. The dropouts incurred caused a total downtime of 12.8 and 17.1 days respectively. Although these estimates are likely conservative (loss of lock occurs for a variety of reasons), they provide an order of magnitude estimate of the potential gains to be made with an early warning system assuming that the incurred downtime could be reduced with sufficient advance notice of the earthquakes' arrivals.

In this paper, we will describe the benefits of a potential early warning system to astronomical observatories. We quantify the ground motion from earthquakes that these detectors are likely to experience and attempt to quantify their effects based on seismic data from near the observatories. 
In section~\ref{sec:Earthquakes}, we outline the background of earthquakes and their identification. 
In section~\ref{sec:hazard}, we describe seismic hazard formalism and their application to observatories. 
In section~\ref{sec:EEW}, we outline the benefits of potential early warning systems to observatories.
In section~\ref{sec:Optimization}, we determine the optimal distance at which a circular seismic array could be deployed to maximize the efficacy of an early earthquake warning network for a telescope in Chile.
In section~\ref{sec:conclusions}, we offer concluding remarks and suggest directions for future research.

\section{Earthquake Properties and Identification}
\label{sec:Earthquakes}

We begin by presenting a review of earthquakes and their identification. Seismic waves can be split into two major categories, surface and body waves. Surface waves travel across the surface of the Earth, while body waves travel through the interior. Body waves propagate in three dimensions, radiating away from the epicenter. Surface waves instead propagate in two dimensions, which means they decay more slowly with distance than the body waves. Surface waves also tend to have larger displacement amplitude than body waves, which increases the damage they cause. As body waves travel through the Earth, they are refracted by variations in seismic speed.

There are two predominant types of body waves, both of which will be important for low-latency earthquake identification. The first are primary waves (P-waves), which are compressional longitudinal waves. They are the fastest of the wave types, with typical speeds ranging from 6-13\,km/s (although in some soils these velocities can be much lower), and therefore arrive at seismometers first. Low-latency epicenter location estimates typically rely on the P-wave arrivals at a number of stations. The second category of body waves are secondary waves (S-waves), also known as shear waves, which are transverse waves. Shear waves arrive after the P-waves, with typical speeds ranging from 3-6\,km/s (although, again, in some soils these velocities can be much lower). They do not travel through fluids, such as the outer core, as shear stresses are not supported in fluids. 

Surface waves come in three different flavors, Rayleigh waves, Love waves, and Stonely waves (Stonely waves are interface waves and are not important for this discussion). 
Rayleigh waves travel along the surface of the Earth with a velocity lower than those of the body waves. Love waves also travel along the surface of the Earth, but with velocity equal to that of shear waves. They travel more slowly due to lower shear wave velocities near the surface of the Earth. Love waves also experience significant dispersion as they travel along the surface, which stretches out their arrival at a given location in time. They also tend to have larger amplitudes than body waves. 
Large-amplitude earthquakes are capable of driving up the normal modes of the Earth \cite{MoRo2008}. Earth's slowest normal-mode oscillation occurs at about 0.3\,mHz, and distinct modes can still be identified up to a few millihertz. At higher frequencies, the discrete vibrational spectrum transforms into a quasi-continuous spectrum of seismic vibrations that are increasingly driven by local sources.

\begin{figure*}[t]
\hspace*{-0.5cm}
 \includegraphics[width=5in]{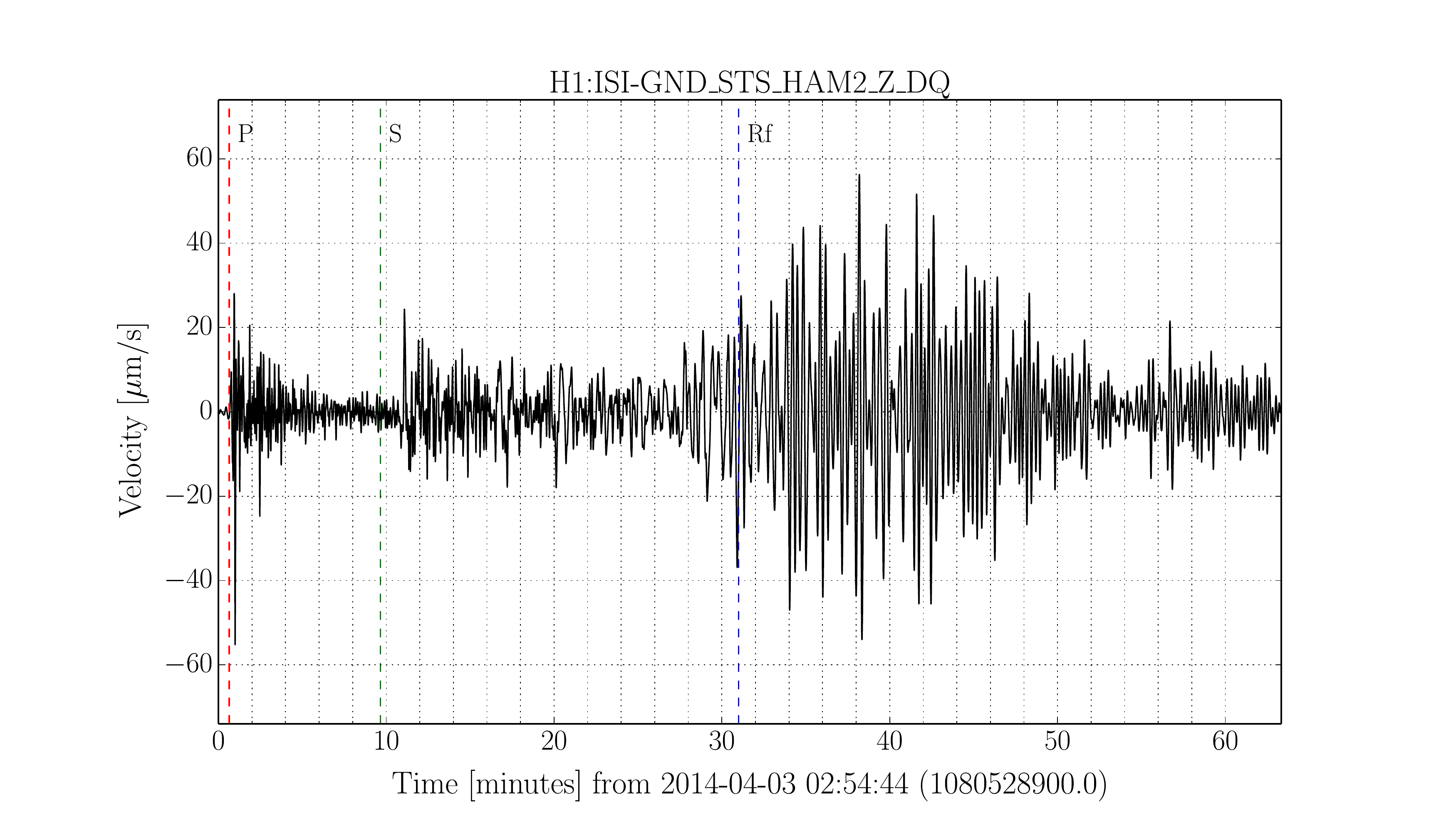}
 \caption{Time-series of vertical ground motion from the April 1, 2014, 8.2 magnitude earthquake near Iquique, Chile at the LIGO Hanford site located in Washington, USA. The P, S, and surface waves all have distinctive arrivals and are indicated with P, S, and Rf respectively. The surface waves are shown to last tens of minutes.}
 \label{fig:timeseries}
\end{figure*}

We now briefly describe the process by which earthquakes are identified and what the ground motions due to the events look like. Figure~\ref{fig:timeseries} shows example timeseries at a seismic station at LIGO Hanford for the April 1, 2014, 8.2 magnitude earthquake near Iquique, Chile. After an earthquake occurs, seismometers nearest the epicenter first record the P-wave arrival times. Using a number of arrival times, the time, latitude, longitude, and depth of the event can be determined. Magnitude estimates come later, which measure the total moment release of the earthquake. This is computed by multiplying the distance a fault moved and the force required to move it. This corresponds to the product of the fault area, the average displacement, and the rigidity modulus. Seismometers read out the ground velocity beneath them. Conventional seismometers are simple to convert from raw data to ground velocity, as their frequency response is flat between 0.008\,Hz -- 70\,Hz and relative calibration errors of broadband seismometers lie well below 0.1 \cite{VeEA2009}. 

At the point in time when latitude, longitude, and depth information for an earthquake have been determined, predictions of the earthquake's effect on places away from the epicenter can be made. In general, both P- and S-wave arrival times, which depend only on the distance between the epicenter and location in question, can be accurately determined. Using the iaspei-tau package \cite{IaspeiTau} wrapped by Obspy \cite{ObsPy}, travel times for the P (pressure) and S (shear) wave components are calculated for known earthquakes. Approximate arrival times for the surface waves are calculated assuming constant 3.5\,km/s speed values. 

\section{Hazard Assessment}
\label{sec:hazard}

We have outlined the properties of earthquakes, and we now turn our attention to the effect of earthquakes on the observatories. To do so, we analyze the effect that seismic ground motion has on observatories. We then examine historical earthquake catalogs and predict the likely ground motion. We then use seismic data from on site observations to predict how ground motion will affect the observatories.

\begin{figure*}[t]
\hspace*{-0.5cm}
 \includegraphics[width=2.9in]{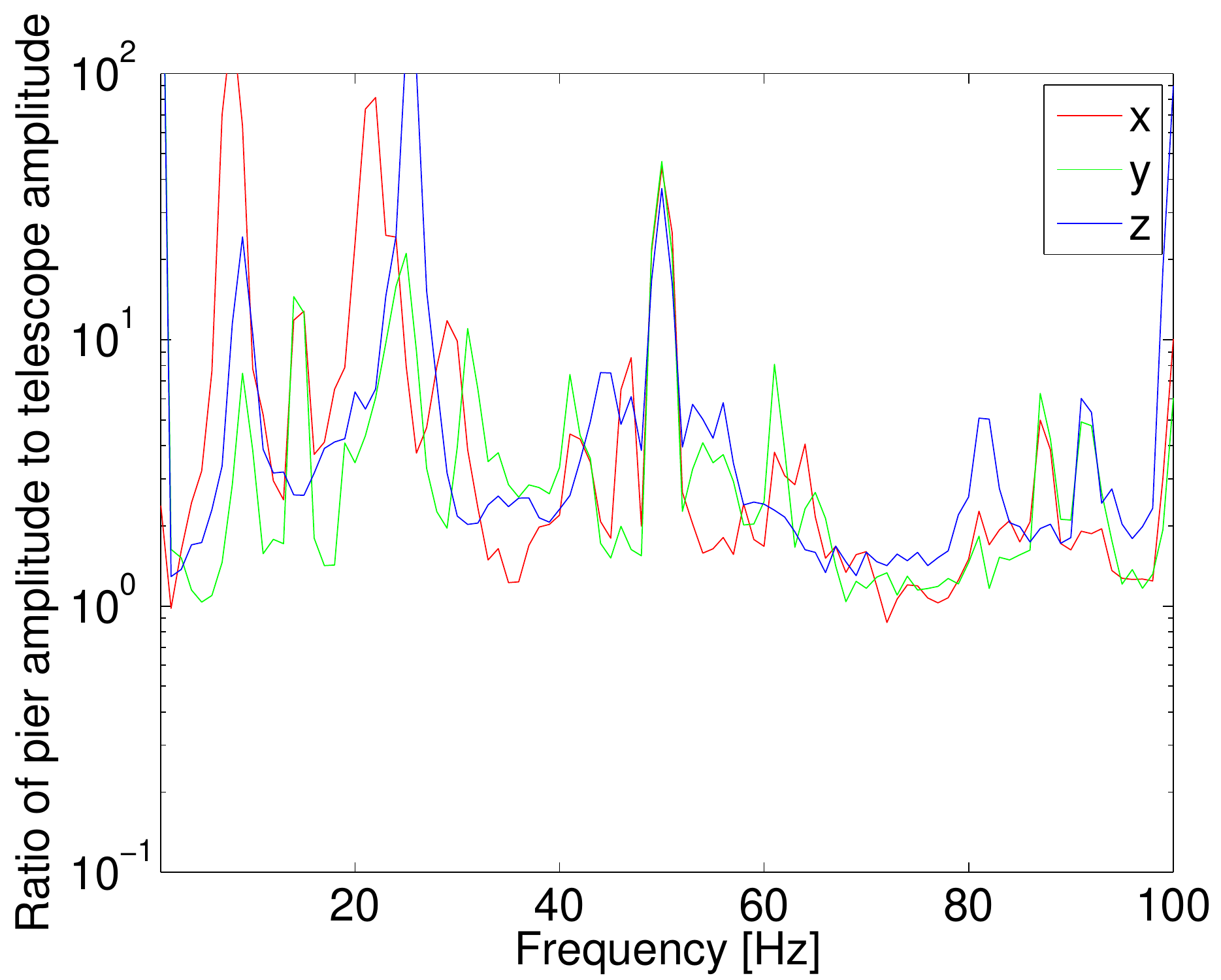}
 \includegraphics[width=2.9in]{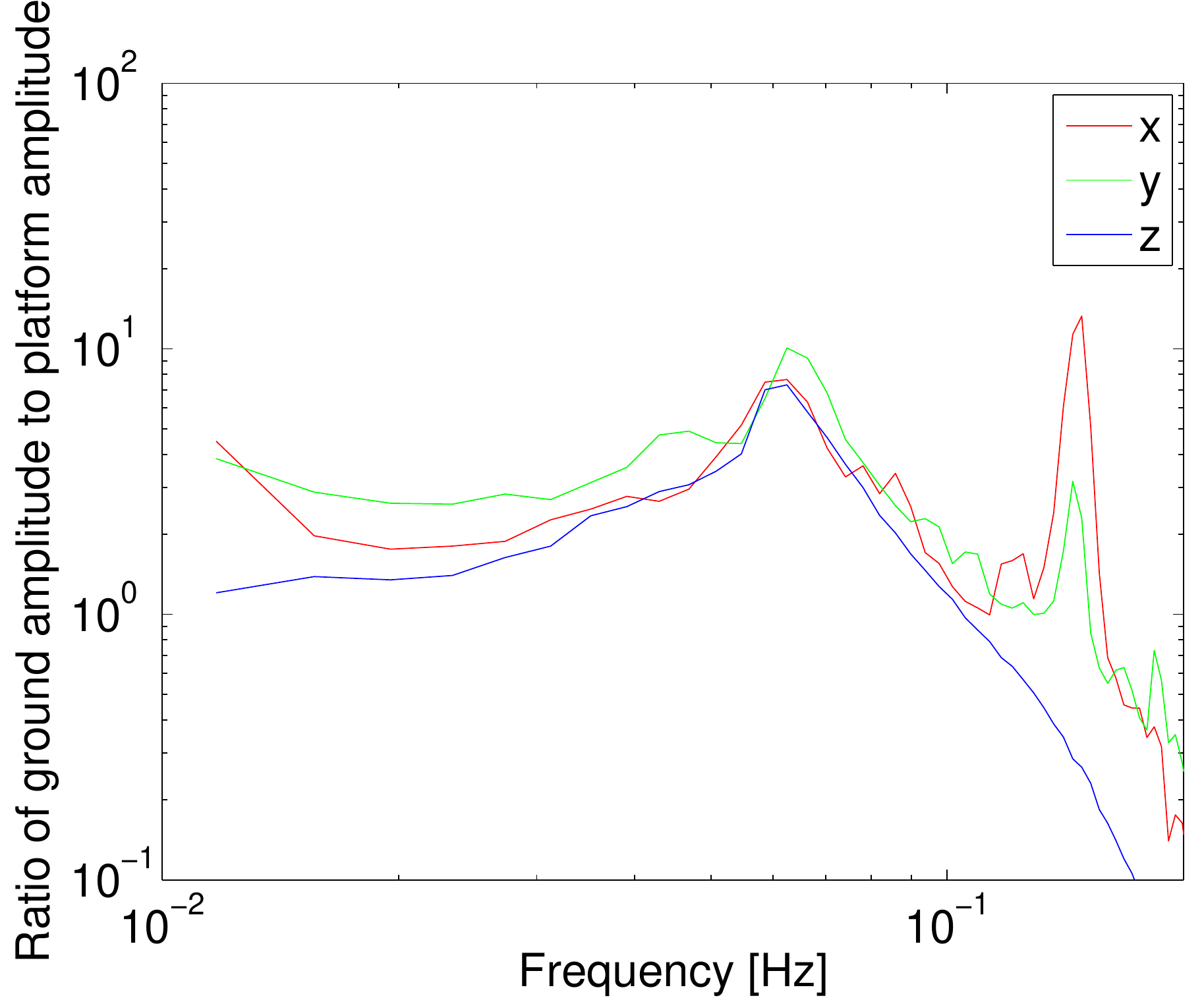}
 \caption{Effect of strong ground motion on the observatories. The plots show the ground motion transfer function for both SOAR and LIGO observatories. SOAR has resonances at 7.5, 20, and 29\,Hz, resulting in over an order of magnitude amplification of the ground motion at those frequencies. LIGO also has significant amplification at earthquake frequencies below 1\,Hz, particularly at 0.06 and 0.15\,Hz. 
 }
 \label{fig:transfer}
\end{figure*}

We now determine a transfer function which we can be used to scale the seismic ground motion to the effect on the astronomical observatories. Due to the coupling of ground motion to these systems, the transfer function between the ground motion and observatory instruments can be significantly greater than one. While telescopes often contain non-linear structures, including base isolators, viscous dampers, seismic restraints, or seismic releases \cite{KaAn2008}, while gravitational-wave detectors can experience significant upconversion \cite{S6DetectorChar}, the transfer functions calculated here are approximations in that linear coupling is assumed. One can compute the transfer function by approximating the system as a series of oscillators with varying natural frequency where each oscillator undergoes the same ground motion. This model can be used to model the response of the structure assuming both the spectrum of the ground motion, the natural frequency of the structure, and the seismic speeds of the ground around the telescope is known. Instead, we determine empirical transfer functions using seismic stations at the observatories during large scale ground motion, one on the ground and another on the instrument platforms. As an example telescope, we use the Southern Astrophysical Research (SOAR) telescope, which is a 4.1\,m optical and near-infrared telescope located on Cerro Pachón, Chile. As example gravitational-wave detectors, we use the LIGO gravitational-wave detector located at Hanford, WA (H1) and Livingston, LA (L1). We use seismometers on the secondary mirror cage in the case of SOAR and an optical platform in the case of the gravitational-wave detector. These plots can be seen in figure~\ref{fig:transfer}. SOAR has resonances at 7.5, 20, and 29\,Hz, resulting in over an order of magnitude amplification of the ground motion at those frequencies. LIGO also has significant amplification at earthquake frequencies below 1\,Hz, particularly at 0.06 and 0.15\,Hz. The idea is that we can multiply the spectrum of the local ground motion by this function to find the response motion of the observatories. We see that both observatories suffer from order of magnitude amplifications of ground motion. This verifies our assumption from above that even 0.1\,g earthquakes can induce significant accelerations for the telescopes, and relatively minor ground motion at gravitational-wave detectors is sufficient to create significant motion at the test masses. 

We now turn our attention to predicting seismic ground motion based on earthquake parameters. 
Probabilistic seismic hazard analysis (PSHA) is a well-developed field \cite{Co1968}. In the following, we will concentrate on the application of telescopes in Chile. To be specific, we take the Magellan telescopes, which are a pair of 6.5\,m diameter optical telescopes located at Las Campanas Observatory in Chile.
The basic goal is to predict expected ground motions given two components.
The first is a distribution of potential earthquake magnitudes and locations. This can be generated from an historical earthquake record. The second is a ground motion prediction model, which describes the probability distribution of local ground motion intensity as a function of many predictor variables such as magnitude, distance, faulting mechanism, the near-surface site conditions, the potential presence of directivity effects, and others \cite{Ba2008}. These models are usually developed using statistical regression on observations from large libraries of observed ground motion intensities and are known to have large variability. The variability is due to prediction of a highly complex phenomenon (ground shaking intensity at a site) using very simplified predictive parameters such as magnitude, distance, and the others described above. We attempt to represent earthquake rupture, which is a complex spatial-temporal process, as a single number such as magnitude, which measures the total seismic energy released in the variable-slip rupture. Additionally, a metric such as distance ignores the complex non-linear wave scattering and propagation through the Earth that seismic waves undergo. And although more detailed models may improve the predictions, more detailed predictions of future earthquakes would also require more knowledge than magnitude and distance probability distributions. Therefore, the significant uncertainties associated with ground motion prediction is an inherent variability in the earthquake hazard environment that must be accounted for when identifying a design ground motion intensity.

We perform a Monte Carlo simulation using information from the historical earthquake record to determine the properties of earthquakes that we expect could harm them. This simulation requires the expected peak ground acceleration (PGA) given earthquake magnitude and distance, PGA(R,M), the probability of having an earthquake of a certain size, $P(M_i = m_j)$, and the probability of an earthquake occuring a certain distance away, $P(R_i = r_k)$.

\begin{figure*}[t]
\hspace*{-0.5cm}
 \includegraphics[width=2.9in]{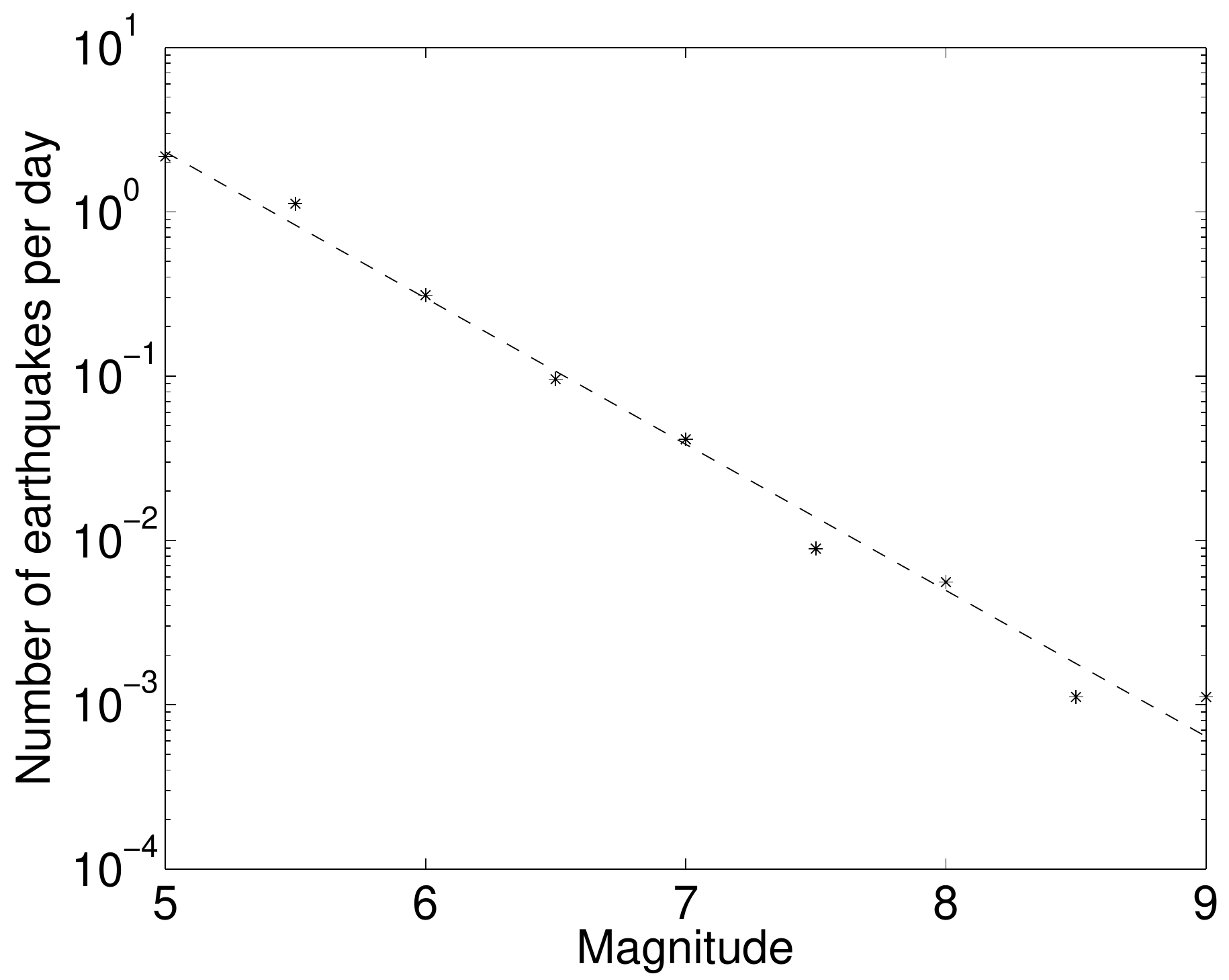}
 \includegraphics[width=2.9in]{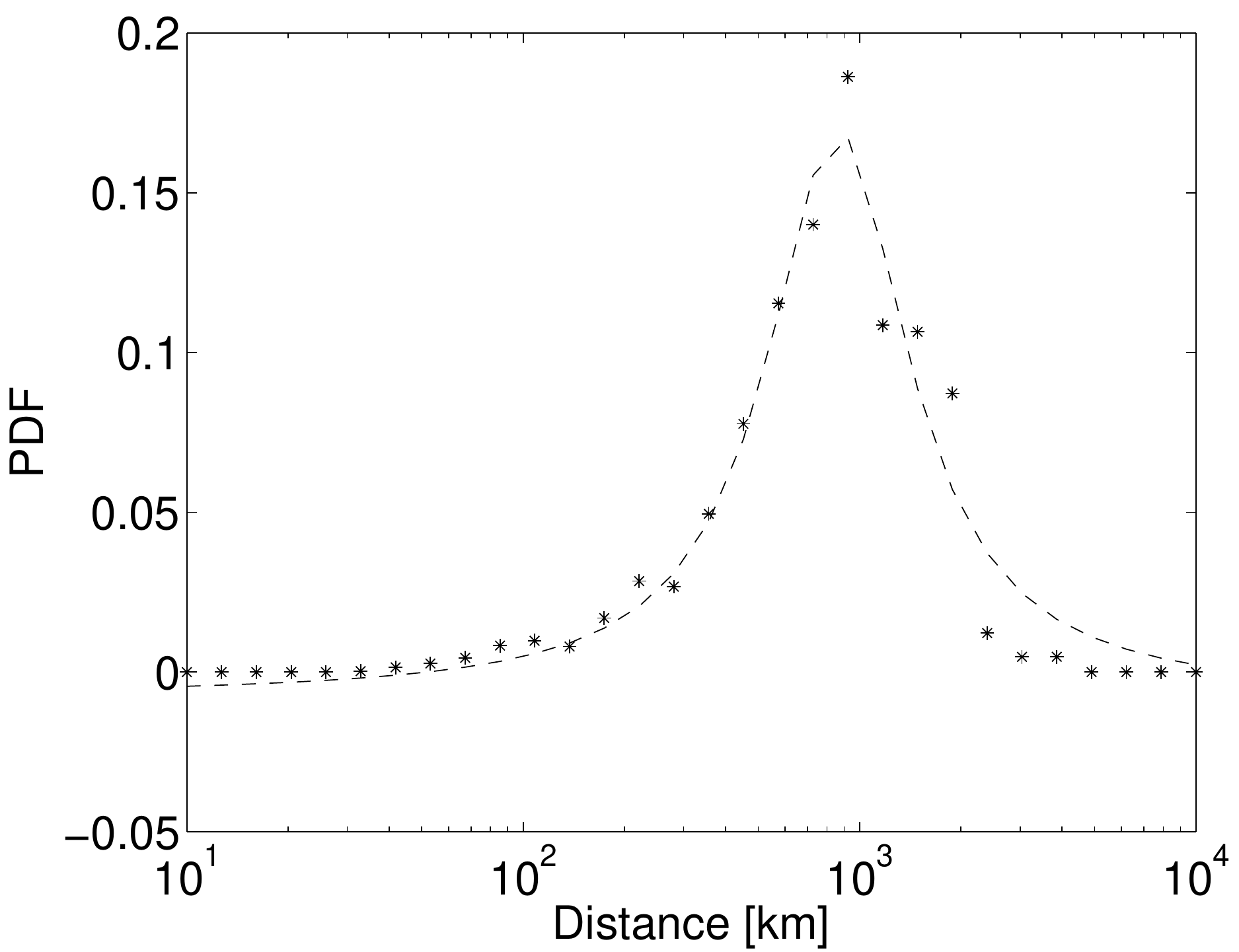}
 \caption{The left plot shows the magnitude distribution of regional Chilean earthquakes and a log fit to the distribution. Earthquakes with a source magnitude $m \geq 5$ occur at a rate of about 3.7 per day in the region between latitudes of $-80 \leq \psi \leq -10$ and longitudes of $-80 \leq \lambda \leq -60$. The right plots shows the distribution of distances between the same earthquakes and the Magellan telescope and a Lorentzian fit to the distribution. The cutoff at large distances is due to the requirement that earthquakes are regional. The cutoff at short distances is due to the lack of large earthquakes that have occurred within a few kilometers of Magellan.}
 \label{fig:fits}
\end{figure*}

The probability distribution of earthquake magnitudes was first studied by Gutenberg and Richter \cite{GuRi1944}, who noted that the distribution of these earthquake sizes in a region generally follows a particular distribution, given as follows
\begin{equation}
\textrm{ln}(\lambda_M) = a + b \times M
\label{eq:GR}
\end{equation}
where $\lambda_M$ is the rate of earthquakes with magnitudes $M$, and $a$ and $b$ are constants. This equation is called the Gutenberg-Richter recurrence law. The $a$ and $b$ constants from equation~\ref{eq:GR} are typically estimated using statistical analysis of historical observations, with additional constraining data provided by other types of geological evidence. The $a$ value indicates the overall rate of earthquakes in a region, and the $b$ value indicates the relative ratio of small to large magnitudes (typical $b$ values are approximately equal to -1). From this equation, we can compute the magnitude distribution of earthquakes that are larger than some minimum magnitude $M_{\textrm{min}}$.
\begin{equation}
f_M (M) = b \textrm{ln}(10) \frac{10^{-b ( M - M_{\textrm{min}})}}{1 - 10^{-b ( M_{\textrm{max}} - M_{\textrm{min}})}}
\end{equation}
We can measure this probability density function empirically using data from regional Chilean earthquakes. The plot on the left of figure~\ref{fig:fits} corresponds to the probability density function of the magnitude distribution of regional Chilean earthquakes from the past forty years, which we approximated as being between latitudes of $-80 \leq \psi \leq -10$ and longitudes of $-80 \leq \lambda \leq -60$, using a bin size of 0.5 in magnitude. The probability density function is fit to a semi-log scale consistent with Eq.~\ref{eq:GR}. 

The second factor we must calculate is $P(R_i = r_k)$, which corresponds to the probability density function of the distance of the earthquakes to the site of interest. This can be computed analytically for a number of fault types. This captures the spatial earthquake density about the facility of interest and depends on the local fault and subduction zone structure. In our case, we can use the same set of historical earthquakes in Chile to empirically calculate $P(R_i = r_k)$. 
On the right is the probability density function of the distance from the Magellan telescope, to which we have fit a Lorentzian distribution. 
We have taken all earthquakes with magnitude $\geq 5$ in this study. We use this cut as earthquakes with magnitudes smaller than this value need to be $\leq 10$\,km from the telescope to create significant ground motion.

The PGA(R,M) is a function of both magnitude and distance to the epicenter. There are specific equations used for Chile. Medina \cite{Med1998} gives the horizontal peak accelerations due to shallow interplate earthquakes (plate interface) as
\begin{equation}
\textrm{PGA} = 733 e^{0.7 M} (R + 60)^{-1.31},
\end{equation}
while Ruiz and Saragoni \cite{RuSa2005} give an expression for intermediate depth earthquakes (H $\geq$ 70\,km)
\begin{equation}
\textrm{PGA} = 565898 e^{1.29 M} (R + 80)^{-3.24},
\end{equation}
where PGA is in $\textrm{cm}/\textrm{s}^2$, M is magnitude, and R is the distance to the earthquake epicenter in km. 

\begin{figure*}[t]
\hspace*{-0.5cm}
 \includegraphics[width=2.9in]{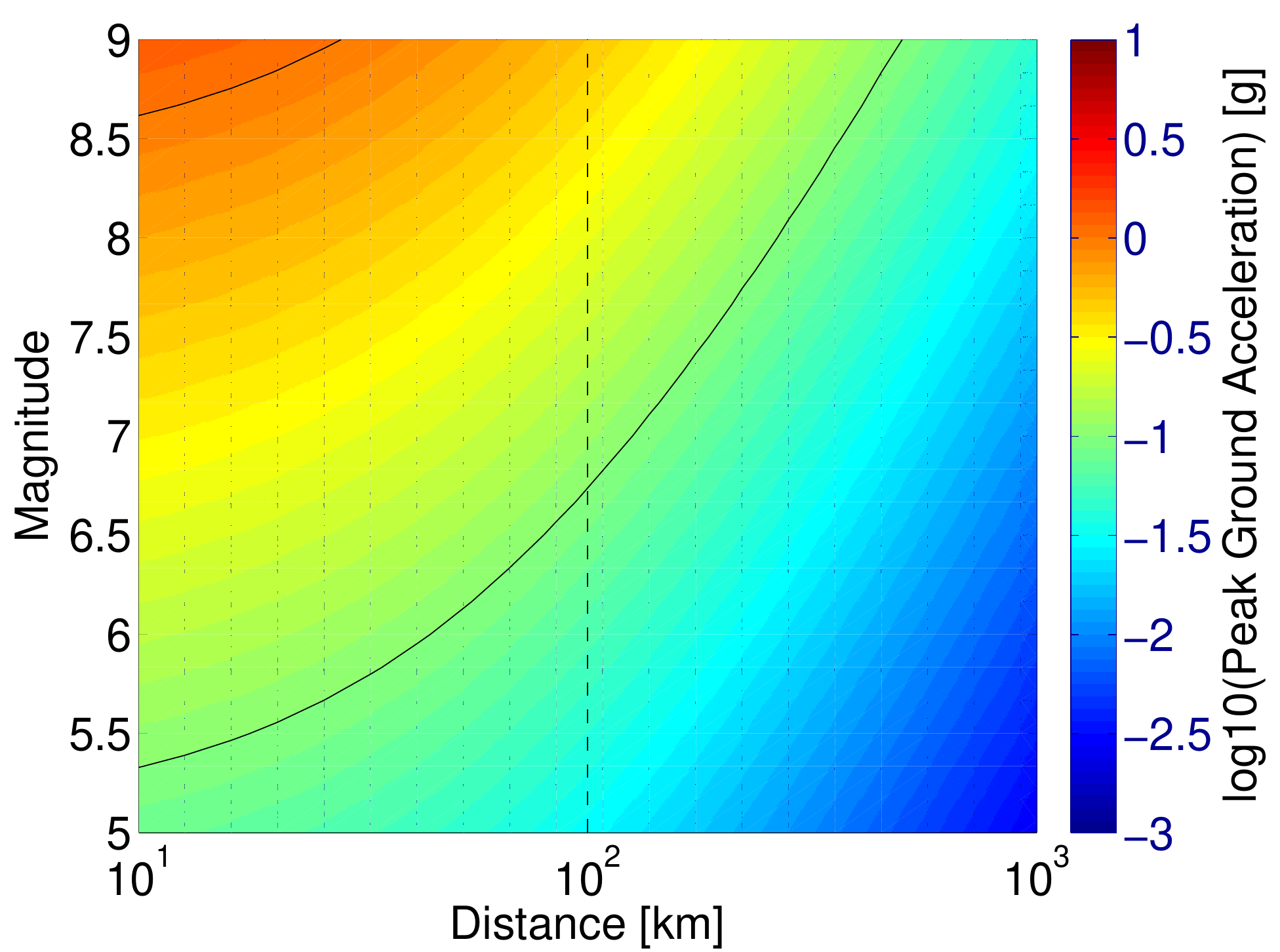}
 \includegraphics[width=2.9in]{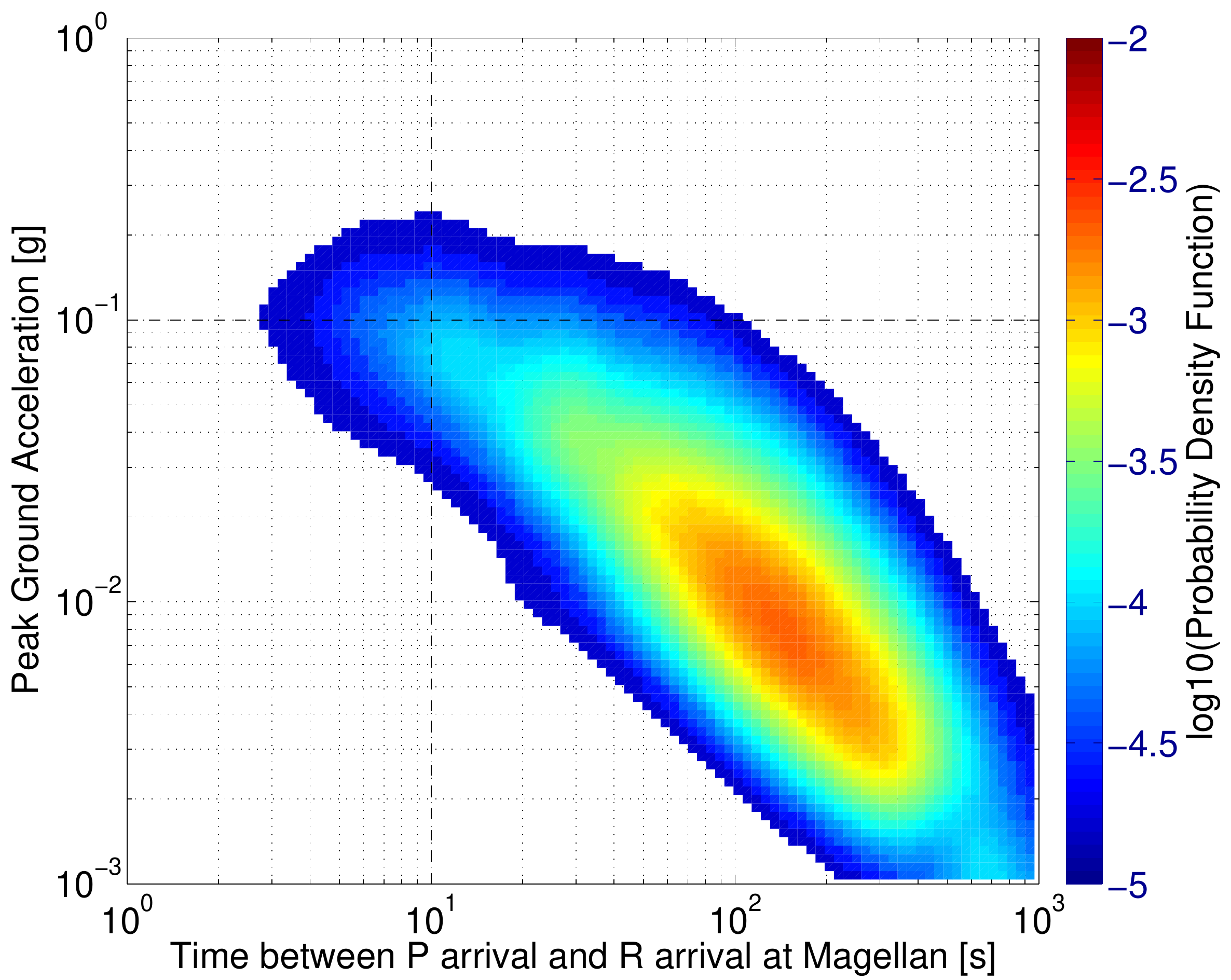}
 \caption{On the left is the predicted peak ground acceleration as a function of magnitude and distance. This used the Medina \cite{Med1998} model for the distribution of ground motion intensity. The black solid lines denote the 0.1\,g and 1\,g of ground motion predictions. The black dotted line corresponds to earthquakes at 100\,km, which provide about 10\,s of warning time for an on-site seismometer. On the right is a Monte Carlo simulation of regional Chilean earthquakes based on the historical record. We denote with dotted lines the earthquakes that correspond to 10\,s of warning time and 0.1\,g of peak ground acceleration.
 }
 \label{fig:PSHA}
\end{figure*}

The left panel of figure~\ref{fig:PSHA} shows the model for peak ground acceleration as a function of magnitude and distance for the Medina \cite{Med1998} model using the historical earthquake record. Based on the above equations, we expect that shallow earthquakes, corresponding to depths $\leq70$\,km, are predicted to have the greatest effect on the telescopes. The plot shows that 0.1\,g peak ground acceleration can occur for earthquakes magnitude 6.5 and 100\,km away. As shown below, telescope resonances can amplify this ground motion significantly. Therefore, taking steps to limit their effects on the detectors will be useful. On the right is a Monte Carlo simulation of regional Chilean earthquakes using the Medina model as well (we found that shallow earthquakes are much more common and have larger amplitude, and so this assumption is reasonable). We denote with dotted lines the earthquakes that correspond to 10\,s of warning time and 0.1\,g of peak ground acceleration. These correspond to a distance of 100\,km with depths $\leq70$\,km. We assume that about 10\,s warning time will be required to make a positive impact in terms of telescope safety. We see that a significant number of events with warning time greater than 5\,s will create significant ground motion. The time axis on the plot corresponds to warning time at Magellan assuming a seismometer on site (which is the difference in P and surface wave arrivals). 

We can also calculate based on these results the annual rate of exceedance of a given ground motion intensity. About 1\% of earthquakes with magnitude $\geq6.0$ cause ground motions at Magellan exceeding 0.1\,g of peak ground acceleration (including magnitudes smaller than this does not change the results sigificantly). There are about 7.8 earthquakes with magnitude $\geq6.0$ per year in Chile. Therefore, there is about an 8\% chance of exceeding 0.1\,g of peak ground acceleration per year. 

\section{Early Earthquake Warning}
\label{sec:EEW}

We now explore the potential increase of warning times (above and beyond using the difference between surface and P-wave arrivals at a seismometer on site) using seismic networks.
Astronomical observatories such as LIGO and LSST have or will have seismometers on site, which are capable of exploiting the difference in P to surface wave arrival differences. Below, we explore the benefits of augmenting these seismometers with off-site sensors.
As an example application, we take the Magellan telescopes as well as the LIGO gravitational-wave detectors.
We used the historical earthquake record and propagation model to compute the time between an event and the arrival of the surface waves at these observatories. In the case of Magellan, we use earthquakes from the Monte Carlo analysis in section~\ref{sec:hazard} which exceed $a \geq 0.1$\,g.
We use global earthquakes for the LIGO sites. 
This difference dramatically affects the timescales.

\begin{figure*}[t]
\hspace*{-0.5cm}
 \includegraphics[width=3.2in]{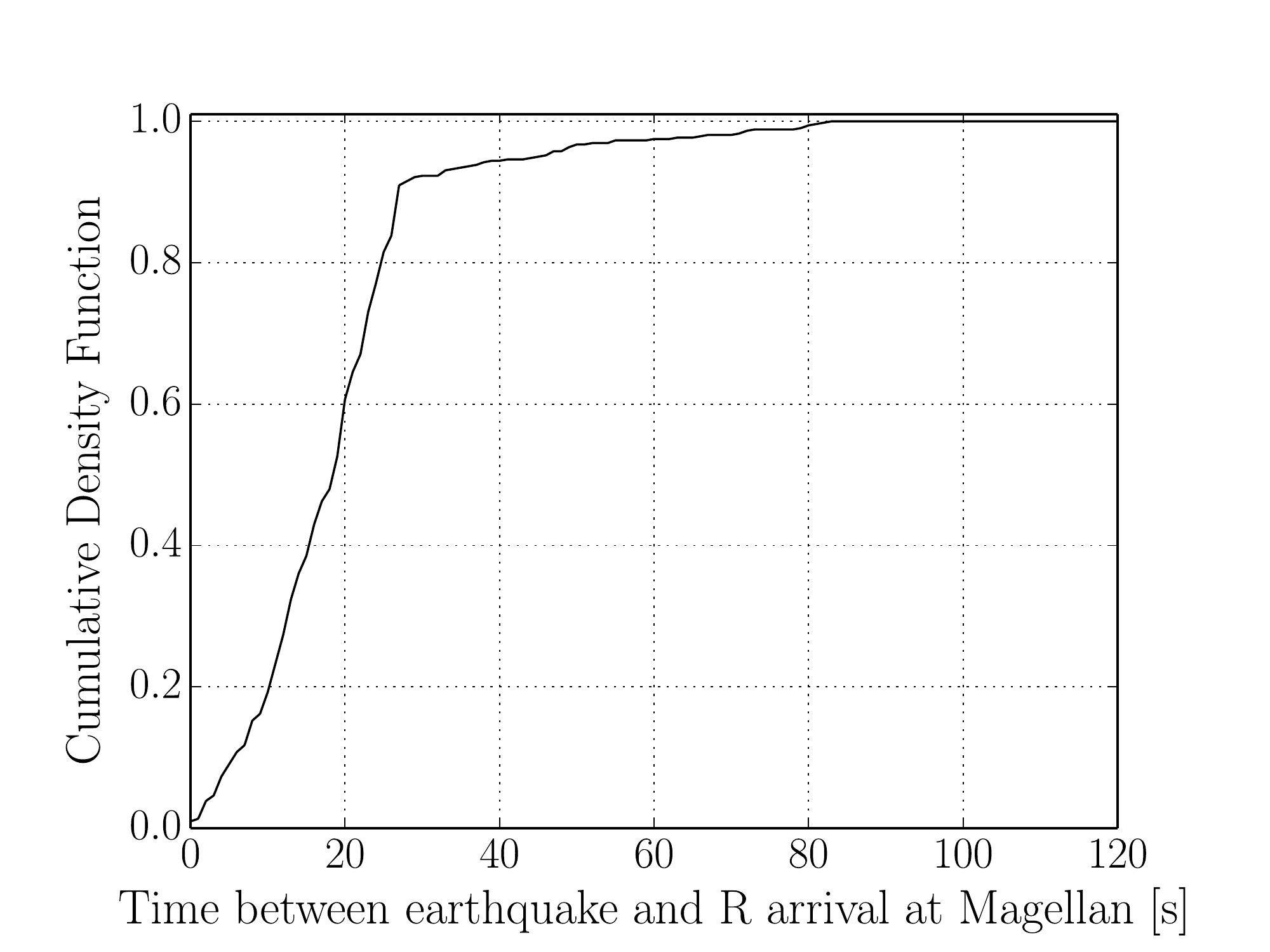}
 \includegraphics[width=3.2in]{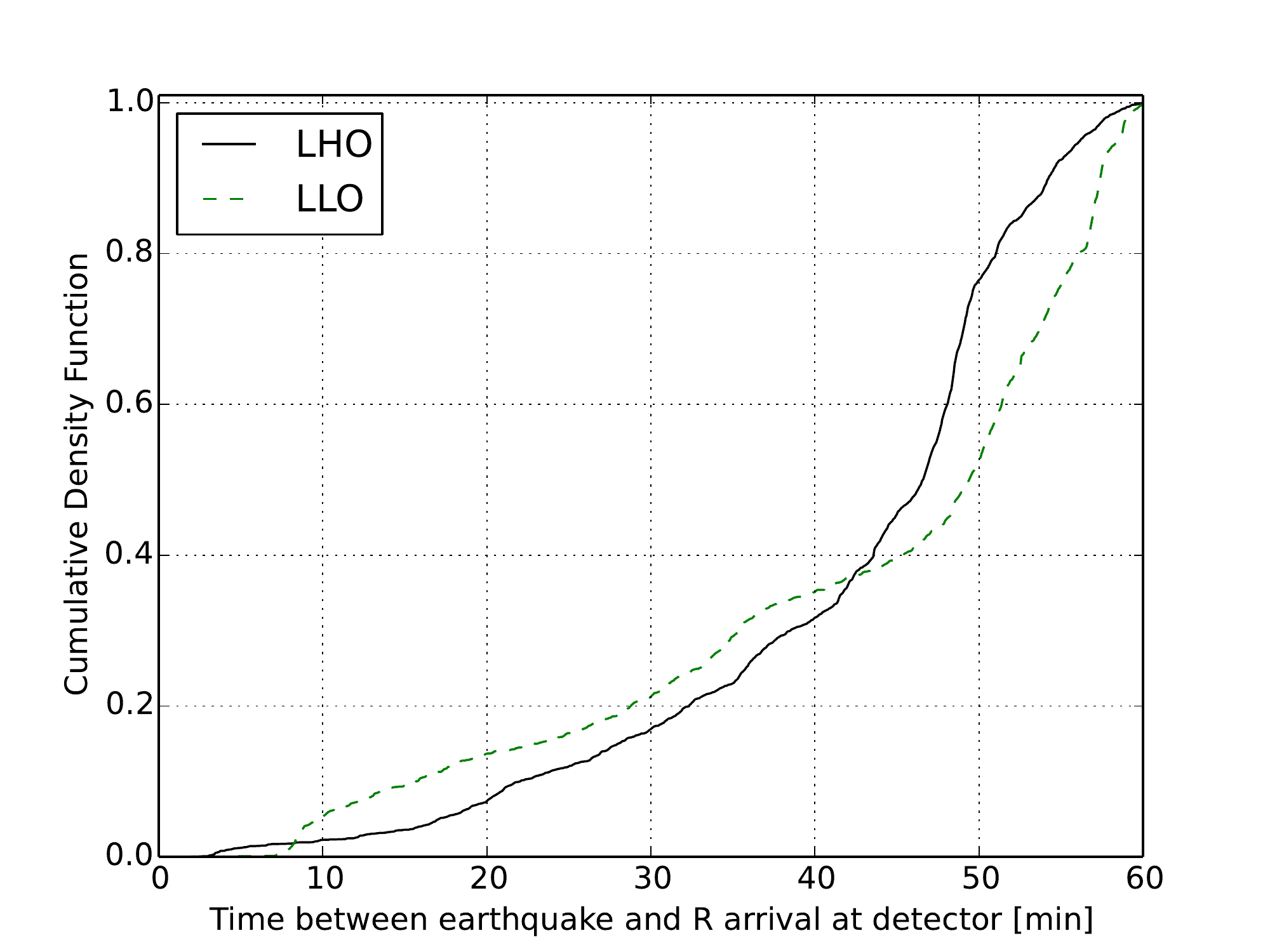}
 \caption{The left plot shows the time delay between the earthquake and approximate arrival of surface waves at Magellan from the Monte Carlo analysis in section~\ref{sec:hazard} which exceed $g \geq 0.1$.
 A majority of the earthquakes are at distances such that their surface waves arrive within 1 minute to the site. The right shows the same at the LIGO Hanford and Livingston sites for global earthquakes (as LIGO is susceptible to all earthquakes). A majority of the locations allow for more than 10 minutes of time between earthquake and site arrival. These delays are optimistic in the sense that an earthquake warning requires a P-wave arrival at at least one seismic station.}
 \label{fig:delays}
\end{figure*}

Figure~\ref{fig:delays} shows the cumulative probability distribution of time delays between the earthquake and approximate arrival of surface waves for Magellan and the LIGO Hanford and Livingston sites. For Magellan, the majority of earthquakes have surface waves which will arrive less than 1 minute after the earthquake occurs. For LIGO, the amount of time between earthquake and surface wave arrival is typically more than 10 minutes. 

It is at this point that the setup for EEW systems for telescopes and gravitational-wave detectors diverge. Due to the significant time-delay, gravitational-wave detectors do not require regional seismic networks with low-latency warning. Instead, we can rely on an existing world-wide notification system from the United States Geological Survey (USGS) that uses available seismic networks to create earthquake notices. Location and magnitude estimates for these events are typically generated within a few minutes by USGS and distributed for observatory use through USGS's Product Distribution Layer (PDL) \cite{PDL2012}, which has been configured to receive all notifications of earthquakes worldwide. This portal will be sufficient for the purposes of current gravitational-wave detectors, which are typically built far from subduction zones (the future KAGRA interferometer at the Kamioka mine in Japan could be an exception to this rule \cite{AsEA2013}).

\begin{figure}[t]
\hspace*{-0.5cm}
 \includegraphics[width=7in]{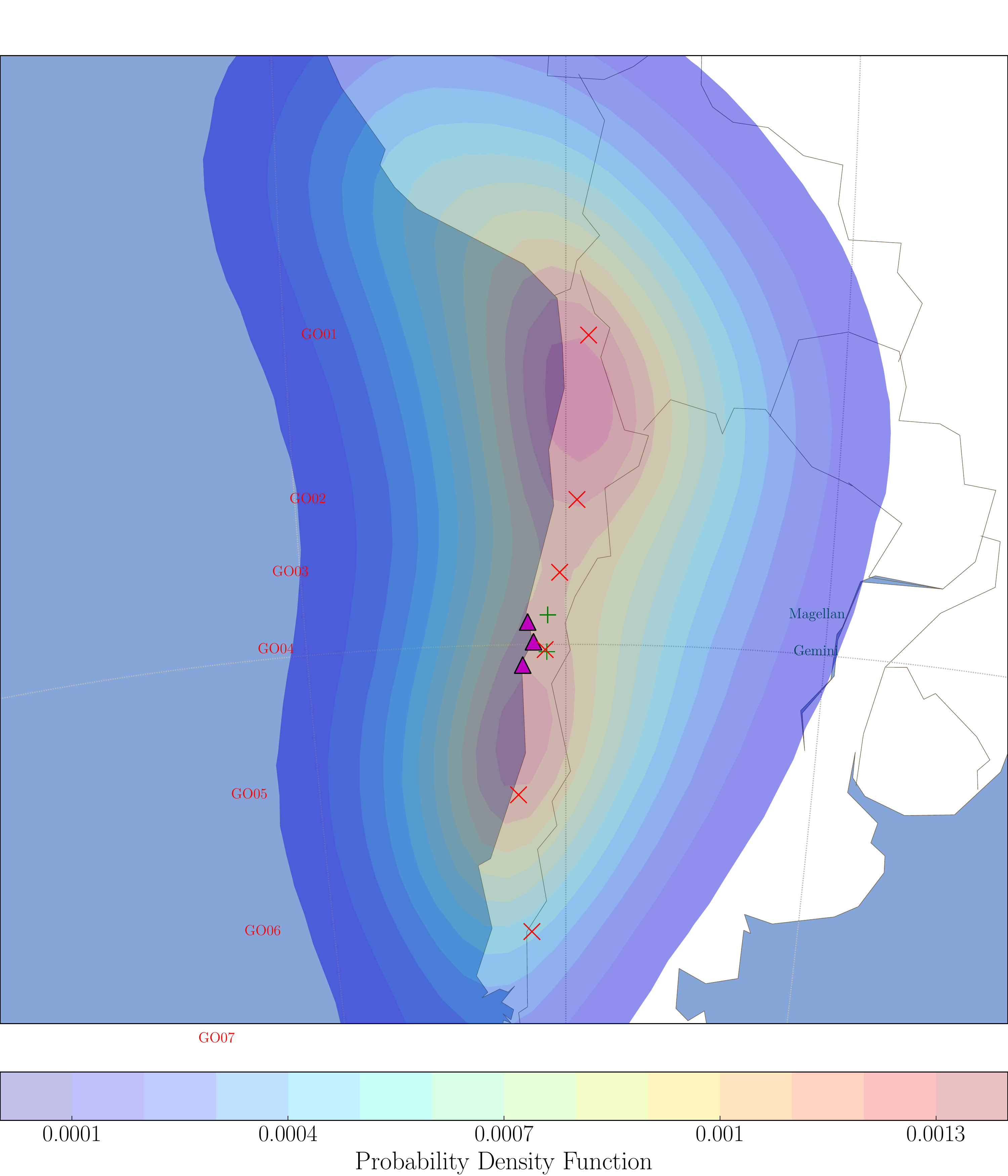}
 \caption{Map of the probability distribution for regional Chilean earthquakes, as well as the locations of current broadband seismometers and telescope locations. The broadband seismometers are represented by red Xs, the telescopes by green crosses, and the proposed low latency alert sensors by purple triangles. We have taken all earthquakes with magnitude $\geq 5$ for this plot.}
 \label{fig:map}
\end{figure}

Due to the short time-delays required for optical and radio telescopes, this system will not be sufficient. Earthquake early warning in a regional setting (as is required for major telescopes) is difficult. The first requirement is a robust data acquisition and transmission system for the seismic data with limited reception delays. The second is identification of the P-waves in seismic sensor timeseries. This is a well-studied problem in seismology and not difficult for the large magnitude of the earthquakes problematic for this analysis \cite{ZhTh2003}. There are, however, three potential difficulties to consider. The first is that the magnitude of earthquakes is difficult to determine quickly. Large earthquakes have energy releases which can last tens of seconds, and so often magnitudes can only be determined after the entire rupture has occurred. Global GPS is often used to determine the overall magnitude of the earthquake by measuring the actual displacements of the Earth's surface close to the fault. Conventional seismometers and accelerometers roll-off in sensitivity at low frequencies, and therefore are not able to provide accurate displacements for this purpose. There is a large effort in developing methodologies that allow magnitude estimation with the initial portion of the P-wave arrival \cite{KuAl2013a}. A third issue is that the communication of seismic signals from the seismic sensors to the location where the data is processed must be robust against both power and telecommunication failure. During major earthquakes, power often goes out, as well as many telephone services. Therefore, techniques like using solar panels or uninterrupted power supplies should be used for the seismic stations, and communication that is independent of internet services, such as RF communication, should be employed. Satellite communication is a potential alternative but also has potential delays.

\begin{figure}[t]
\hspace*{-0.5cm}
 \includegraphics[width=5in]{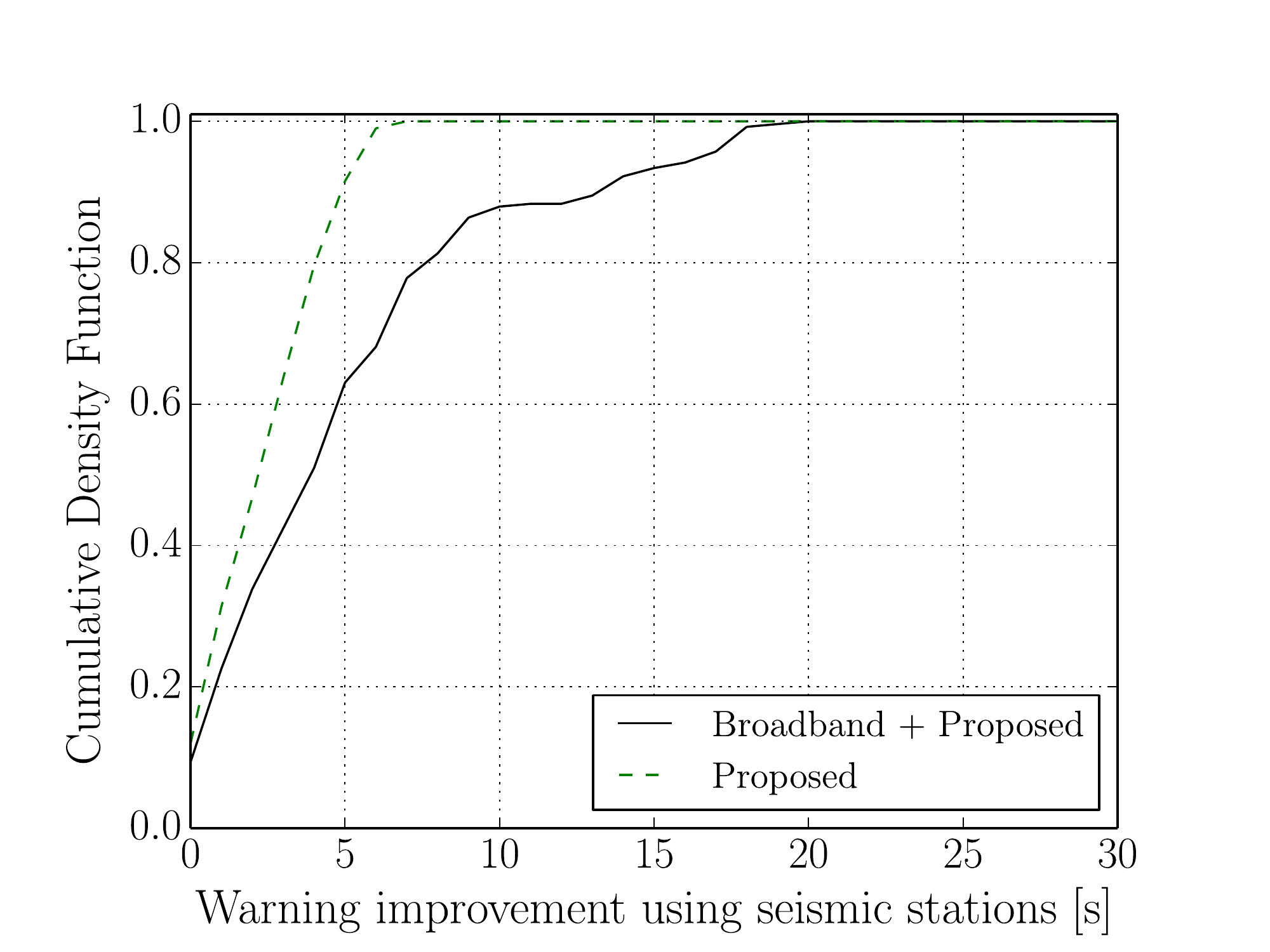}
 \caption{Potential gain in warning time, which is determined by the difference between the P-wave arrival at the nearest seismometer and the surface wave arrival at Magellan, using only the proposed low latency alert network or those sensors in conjunction with the current broadband seismic network. This is the gain on top of using a single on-site sensor. We use earthquakes from the Monte Carlo analysis in section~\ref{sec:hazard} which exceed $g \geq 0.1$. Both show that there is a significant advantage to using a network of some kind over a single on-site sensor.}
 \label{fig:delay}
\end{figure}

Figure~\ref{fig:map} shows a map of the probability distribution for regional Chilean earthquakes, computed using the historical earthquake record, as well as the locations of current broadband seismometers (black Xs), telescope locations (green crosses), and proposed low-latency alert sensors (purple triangles). Currently, there is up to a 5\,s delay in data transmission from the current broadband seismometer stations, which significantly hampers their contribution in a system where every second counts. Therefore, there are three proposed locations for low-latency alert sensors to be placed near the current and future telescope sites. Figure~\ref{fig:delay} shows the potential warning times for a low-latency earthquake network for Chile for earthquakes from the Monte Carlo analysis in section~\ref{sec:hazard} which exceed $a \geq 0.1$\,g. The potential gain in warning time using only the proposed low latency alert network or those sensors in conjunction with the current broadband seismic network is shown. 
These warning times correspond to improvements to the times given by figure~\ref{fig:delays}.
This analysis assumes that the equipment configuration and technology improves such that the 5\,s delay from the broadband seismometers can be significantly reduced. Both show that there is a significant advantage to using a network of some kind over a single on-site sensor to perform the measurement (simply measuring the difference in P-wave and surface-wave arrival at the site). This can be quantified by the expression in Eq.~\ref{eq:deltaTWarning}
\begin{equation}
\Delta t_\textrm{warning} = (t_\textrm{TR} - t_\textrm{SP}) - t_\textrm{latency} = (\frac{\sqrt{r_T^2 + z^2}}{c_R} - \frac{\sqrt{r_S^2 + z^2}}{\alpha}) - t_\textrm{latency}
\label{eq:deltaTWarning}
\end{equation}
where $t_\textrm{warning}$ is the warning time for the telescope, $t_\textrm{TR}$ is the time-of-arrival for the surface waves at the telescopes, $t_\textrm{SP}$ is the time-of-arrival for the P-wave at the nearest alert sensor, and $t_\textrm{latency}$ is the latency associated with taking the seismic trace, determining the presence of a P-wave, and informing the telescope, $r_T$ is the horizontal distance between the earthquake and telescope, $r_S$ is the horizontal distance between the earthquake and nearest alert sensor, z is the earthquake depth, $c_R$ is the speed of surface waves, and $\alpha$ is the speed of P-waves. There are two limiting cases when considering potential warning times. The first is the case that the distance is dominated by horizontal distance. In this case, it is likely that an optimally placed seismic network is able to provide substantial improvements over a warning time based solely on on-site P arrival warning. The second case is that the distance is dominated by vertical separation (i.e. that the earthquake is directly beneath facility). In this case, seismometers on site will provide the only possible warning time, which will correspond to the difference between the P-wave and surface wave arrivals. Thus, the amount of warning time provided will then depend only on the depth of the earthquake.

\section{Seismic network optimization}
\label{sec:Optimization}

We now seek to determine the optimal distance at which a circular seismic array could be deployed to maximize the efficacy of an EEW network. We note here that this analysis is for demonstrative purposes only, please see \cite{StBi2013,PiBi2013,OtBo2010} for applications of genetic algorithms to the problem of optimal sensor placement in EEW networks. We make a number of simplifying assumptions. The first is that P-waves travel at $\alpha=8$\,km/s and surface waves travel at $c_R=3.5$\,km/s. The second is that we are not concerned about false alarms, such that P-wave arrivals at a single sensor can be used to discriminate between events above and below threshold. Because large earthquakes fault over large regions and can take minutes to complete, early and accurate estimation of earthquake magnitudes is nearly impossible. For our purpose, we do not require extreme accuracy, but by placing the seismometers as near to the telescope as possible, we may limit our ability to determine the magnitudes.

\begin{figure*}[t]
\hspace*{-0.5cm}
 \includegraphics[width=3.5in]{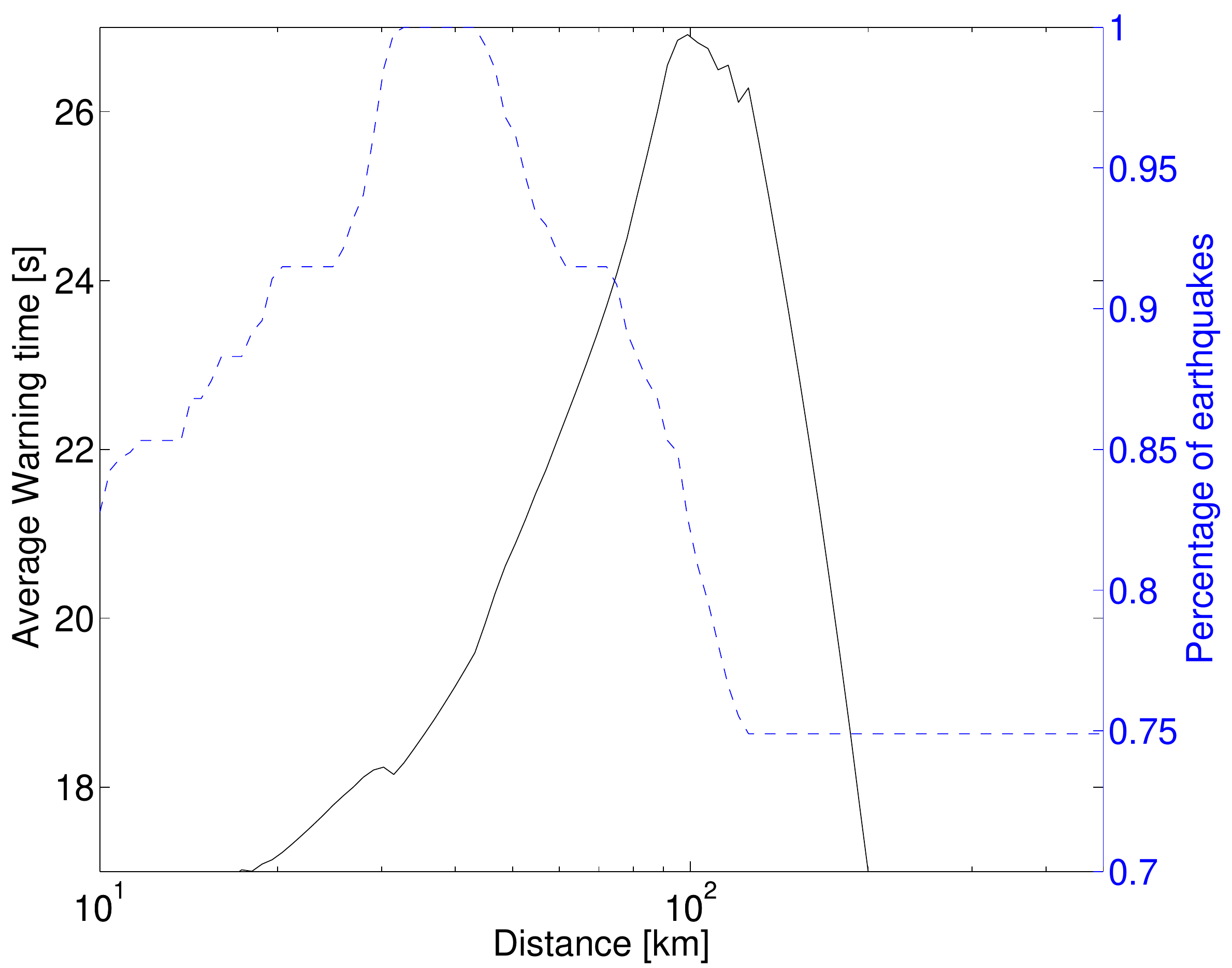}
 \caption{Optimal seismometer distances for the EEW seismic network. The solid line shows the median warning time as a function of distance. The dashed line shows the percentage of earthquakes caught with warning times exceeding 10\,s as a function of distance. The roll-off for larger distances is due to attentuation that suppresses impact of large and distant earthquakes. 
 }
 \label{fig:WarningTimes}
\end{figure*}

We consider two optimization schemes, although more are possible. The first is the case where one desires to catch as many earthquakes as possible (assuming a fixed, minimum warning time of 10\,s). Seismometers on site can give warning times corresponding to the difference in the P- and surface wave seismic velocities, as can be seen by equation~\ref{eq:deltaTWarning}. Assuming $\Delta t_\textrm{warning} = 10$\,s, this gives a minimum earthquake distance of 62\,km. Therefore, improvements to the number of caught earthquakes would require a network at a distance closer than this. To determine this distance, we take the distribution of earthquakes that exceeded the threshold of 0.1\,g from the right of figure~\ref{fig:PSHA} and calculate the warning times for a circular array at fixed distance from the telescope. The dashed line in figure~\ref{fig:WarningTimes} shows that the peak number of earthquakes caught as a function of distance occurs when the sensors are placed between 30-50\,km. Therefore, an EEW seismic system consisting of a circle of seismometers surrounding the telescope would be most usefully placed at that distance to catch the most earthquakes. A second possible optimization scheme is to maximize the warning times for as many earthquakes as possible. The solid line in figure~\ref{fig:WarningTimes} shows that the peak median warning time as a function of distance occurs between 70-100\,km. The peak median warning time is about 25\,s.

These estimates do not account for potential differences in the seismic hazard as a function of distance from the telescope as well as differences in the hazard in different directions from the telescope. An improvement could account for the fact that the lower the magnitude of the earthquake, the more time is required to determine whether it is a large enough of an event to warrant telescope response. Higher magnitude events can be more clearly defined as such based on the P-wave arrival and thus require less analysis time. For the magnitudes of the earthquakes discussed here, this effect is likely negligible, as all P-wave arrivals should be sufficient. More important may be to account for the spatial distribution of earthquakes as given by figure~\ref{fig:map}. If a limited number of seismometers are to be used in an EEW network, due to limited equipment, it could be important to place seismometers in directions that will maximize the warning time from potential earthquakes.

\section{Conclusion}
\label{sec:conclusions}

In this paper, we have discussed the problem of earthquakes for large-scale astronomical observatories. We have shown how it is important to develop an EEW network to maximize the warning time for these experiments. We have discussed how a strong motion local seismic network in Chile could provide more than 10\,s of warning time for conventional telescopes. We have shown that a circular network should be placed nearer than 100\,km away from the telescope. We have also shown that existing worldwide networks of seismometers are adequate for a potential gravitational-wave detector warning system. An earthquake warning system can do two things: predict likely earthquake arrival times and expected minimum ground velocity amplitudes. The earthquake arrival times are required to provide the amount of delay before surface waves arrive, which can facilitate protective action. Approximate ground velocity amplitudes are necessary to determine the type of protective action to take, if at all. Prediction of the ground velocity amplitude based on earthquake magnitude and distance will be required to limit the false alarms. This prediction should account for physical effects with variable parameters used to fit to the seismic data currently available.

The hope is that with knowledge that seismic waves of significant amplitude were about to arrive, scientists on site could take preventative measures to limit the amount of damage and downtime the detectors experienced. In the case of telescopes, most likely this action will be to close mirror covers, telescope dome, and stop slewing of the telescope. For gravitational-wave detectors, this will most likely be a change in gain control for the seismic isolation system. In the future, we hope to provide more detailed models of the response of current and future telescopes to significant ground motion. We intend to implement a gravitational-wave detector early warning system and study the steps that can be taken to minimize the effect of ground motion on the detectors. We will also study potential locations and methods for sensors for a Chilean telescope EEW system.

We have discussed some of the scientific systems that could benefit from early seismic warning. There are other benefits in personnel safety, as well as public utilities, including reactors, and public transportation systems. There are a number of common sense recommendations to be made going forward:
\begin{enumerate}
\item Science facilities: stay abreast of rapidly changing seismic warning systems and make provisions for actions to take in response to seismic warning. 
\item Universities: Distribute low-latency alert to subscribing on-campus devices.
\item Manufacturers: Make provisions for safe-mode response to seismic triggers.
\item Data Centers: Store particularly valuable data at redundant (earthquake-uncorrelated) sites and provide means to take data protection steps upon seismic alert.
\end{enumerate}

\section{Acknowledgments}
The authors would like to thank USGS scientists Paul Earle and Michelle Guy for helpful discussions related to Early Earthquake Warning and the PDL client.
MC was supported by the National Science Foundation Graduate Research Fellowship
Program, under NSF grant number DGE 1144152. 
CWS is grateful to the DOE Office 
of Science for their support under award DE-SC0007881.

\bibliographystyle{unsrt}
\bibliography{references}

\end{document}